\documentstyle[prl,aps,multicol]{revtex}

\begin{document}

\title{On the Disentanglement of States}

\author{Tal Mor\thanks{Electrical Engineering, UCLA, Los Angeles, Cal., USA}}

\date{\today}

\maketitle

\begin{abstract}

Disentanglement is the process which transforms a state 
$\rho$ of two subsystems into 
an unentangled state,
while not effecting the reduced density matrices of
each of the two subsystems.
Recently Terno~\cite{Terno98} 
showed that an arbitrary state  
cannot be disentangled into
a {\em tensor product} of its reduced density matrices.
In this letter we present various novel results regarding disentanglement
of states.
Our main result is that there are sets of states which 
cannot be successfuly
disentangled (not even into a separable state). Thus, we prove that
a universal disentangling machine cannot exist.  

\end{abstract}

\begin{multicols}{2}

Entanglement plays an important role in quantum 
physics~\cite{Peres93}.  
Due to its peculiar non-local properties, entanglement is one
of the main pillars of non-classicality. The creation of entanglement
and the destruction of entanglement via general operations are still under
extensive study~\cite{entanglement}.
Here we concentrate
on the process of disentanglement of states.
For simplicity, we concentrate on qubits in this letter, and on the
disentanglement of two subsystems.

Let there be two two-level systems ``X'' and ``Y''.
The state of each such system is called a quantum bit (qubit).
A pure state which is a tensor product of two qubits
can always be written as $|0({\rm X})0({\rm Y})\rangle$ 
by an appropriate choice of basis,
$|0\rangle$ and $|1\rangle $ for each qubit.
For convenience, we drop the index of the subsystem (whenever it is
possible), and order them
so that ``X'' is at the left side.
By an appropriate choice of the basis $|0\rangle$ and $|1\rangle$,
and using the Schmidt decomposition (see \cite{Peres93}),
an entangled pure state of two qubits can always be written as 
$ | \psi \rangle =  \cos \phi |00\rangle + \sin \phi |11\rangle $
or using a density matrix notation $\rho = |\psi\rangle \langle \psi|$
\begin{equation} \rho =  
[  \cos \phi |00\rangle + \sin \phi |11\rangle ]
[  \cos \phi  \langle 00| + \sin \phi  \langle 11|]
\ .
\end{equation}
The reduced density matrix of each of the qubits 
is
$\rho_{\rm X} = 
{\rm Tr}_{\rm Y} [\rho({\rm XY})] $ and 
$\rho_{\rm Y} = 
{\rm Tr}_{\rm X} [\rho({\rm XY})] $.
In the basis used for the Schmidt decomposition the two reduced density
matrices are
\begin{equation}\label{reduced-state}
\rho_{\rm X} = 
\rho_{\rm Y} = 
\left( \begin{array}{cc}
\cos^2\phi & 0   \\
   0      & \sin^2 \phi
\end{array} \right) 
\ . 
\end{equation}

Following Terno~\cite{Terno98} and Fuchs~\cite{Fuchs}
let us provide the following two definitions
(note that the second is an interesting special case of the first):

\noindent {\em Definition}.---
Disentanglement  
is the process that transforms a state of two
(or more) subsystems into an unentangled state  
(in general, a mixture of product states) such
that the reduced density matrices of each of the subsystems are uneffected.

\noindent {\em Definition}.---
Disentanglement into a tensor product state 
is the process that transforms a state of two
(or more) subsystems into a 
tensor product of the two reduced density matrices. 

We noticed that 
according to these definitions, when a successful disentanglement is
applied onto any pure product state, 
the state must be left unmodified. That is,
\begin{equation}\label{pure-ps}
|00\rangle \longrightarrow |00\rangle 
\end{equation}
(in an appropriate basis).
This fact proved very useful in the analysis we report here.

The main goal of this letter is to show that a universal disentangling 
machine cannot exist.
A universal disentangling machine is a machine that could disentangle
any state which is given to it as an input.
In order to prove that such a machine cannot exist, it is enough to find
{\em one} set of states that cannot be disentangled if the data
(regarding which state is used) is not available.

To analyze the process of disentanglement 
consider the following experiment involving two subsystems ``X'' and
``Y'', and a sender who sends {\em both systems} to the receiver
who wishes to disentangle the state of these two subsystems:  
Let the sender (Alice) and the disentangler (Eve)
define a finite set of states $|\psi_i\rangle$;
let Alice choose one of the states at random, and let it be the
input of the disentangling machine designed by Eve.  
Eve does not get from Alice the data
regarding {\em which} of the states Alice chose, so Eve's aim is to design a
machine that will succeed to disentangle any of the possible states
$|\psi_i\rangle$.

In the same sense that an arbitrary state cannot be cloned 
(a universal cloning machine does not exist~\cite{WZ82}),
it was recently shown by Terno~\cite{Terno98} 
that an arbitrary state cannot be
disentangled into a tensor product of its reduced density matrices. 
Note that this novel result of~\cite{Terno98} proves 
that {\em universal disentanglement into product states} is impossible,
and it leaves open the more general question of whether
a {\em universal disentanglement} is impossible (that is,
disentanglement into separable states).

We extend the investigation of the process of disentanglement well beyond
Terno's novel analysis in  several ways.
First, we find a larger class (then the one found by Terno) of
states which cannot be disentangled into product states.
Then, we show that there are non-trivial
sets of states that {\em can} be disentangled.
In particular, 
we present a set of states that cannot be disentangled into tensor
product states, {\em but} can be disentangled into separable states.
Finally, we present our most important result; 
a set of states that {\em cannot be disentangled}.
The existence of such a set of states proves that a universal 
disentangling machine cannot exist.
Using the terminology of~\cite{WZ82} we can say that our letter shows that
{\em a single quantum
can not be disentangled}.

Consider a set of states containing only one state.
Since the state is known, obviously it can be disentangled. 
E.g., it is replaced by the appropriate tensor product state.

We first prove that there are infinitely many sets of
states that {\em cannot} be disentangled into product states.
Our proof here follows from Terno's method,  
with the addition of using the Schmidt decomposition
to analyze a larger class of states.
The most general form of two entangled states can always be presented 
(by an appropriate choice of bases) as:
\begin{eqnarray} \label{the-states} 
|\psi_0 \rangle &=& \cos \phi_0 |00\rangle + \sin \phi_0 |11\rangle
\nonumber \\
|\psi_1 \rangle &=& \cos \phi_1 |0'0'\rangle + \sin \phi_1 |1'1'\rangle
\ .
\end{eqnarray}
To prove that there are states for which disentanglement into
tensor product states is impossible,
let us restrict ourselves to the simpler subclass
\begin{eqnarray}  
|\psi_0 \rangle &=& \cos \phi |00\rangle + \sin \phi |11\rangle
\nonumber \\
|\psi_1 \rangle &=& \cos \phi |0'0'\rangle + \sin \phi |1'1'\rangle
\ .
\end{eqnarray}
There exists some basis  
\begin{equation} 
 |0''\rangle = {1 \choose 0} ;
 |1''\rangle = {0 \choose 1} 
\end{equation}
such that 
the bases vectors $|0\rangle;|1\rangle$ and $|0'\rangle;|1'\rangle$
become   
\begin{equation} 
 |0\rangle = {\cos \theta \choose \sin \theta} ;
 |1\rangle = {\sin \theta \choose -\cos \theta} \ ,
\end{equation}
and     
\begin{equation} 
 |0'\rangle = {\cos \theta \choose -\sin \theta} ;
 |1'\rangle = {\sin \theta \choose \cos \theta} 
\end{equation}
respectively, in that basis.
The states (\ref{the-states}) are now
\begin{eqnarray} 
|\psi_0 \rangle &=&
c_\phi   
{c_\theta \choose s_\theta}
{c_\theta \choose s_\theta}
  +              s_\phi   
{s_\theta \choose - c_\theta}
{s_\theta \choose - c_\theta}
\nonumber \\
|\psi_1 \rangle &=&
c_\phi   
{c_\theta \choose - s_\theta}
{c_\theta \choose - s_\theta}
  +              s_\phi   
{s_\theta \choose c_\theta}
{s_\theta \choose c_\theta}
\ ,
\end{eqnarray}
with $c_\phi \equiv \cos \phi$, etc.
The overlap of the two states is 
${\rm OL}= \langle \psi_0 | \psi_1 \rangle =
\cos^2 2\theta + \sin 2\phi \sin^2 2 \theta$.
The reduced states are given by 
\begin{equation}
\hat{\rho_0} =  {c_\phi}^2  
\left( \begin{array}{cc}
{c_\theta}^2 & c_\theta s_\theta \\
c_\theta s_\theta & {s_\theta}^2 
\end{array} \right) 
 +        {s_\phi}^2  
\left( \begin{array}{cc}
{s_\theta}^2 &  - c_\theta s_\theta \\
 - c_\theta s_\theta & {c_\theta}^2 
\end{array} \right) 
\ , 
\end{equation}
and 
\begin{equation}
\hat{\rho_1} =  {c_\phi}^2  
\left( \begin{array}{cc}
{c_\theta}^2 &  - c_\theta s_\theta \\
 - c_\theta s_\theta & {s_\theta}^2 
\end{array} \right) 
 +        {s_\phi}^2  
\left( \begin{array}{cc}
{s_\theta}^2 &  c_\theta s_\theta \\
 c_\theta s_\theta & {c_\theta}^2 
\end{array} \right) 
\ . 
\end{equation}
Thus, the state after the disentanglement into tensor product states is
$ (\rho_{\rm disent})_0 = \hat{\rho_0} \hat{\rho_0}$ or  
$ (\rho_{\rm disent})_1 = \hat{\rho_1} \hat{\rho_1}$.

The minimal probability of error for distinguishing 
two states~\cite{Helstrom76}  
is given by
$ {\rm PE} = \frac{1}{2} - \frac{1}{4} {\rm Tr}| \rho_0 - \rho_1| $.
For two pure states there is a simpler expression:
$ {\rm PE} = \frac{1}{2} - \frac{1}{2} \sqrt{[1 - OL^2]}$.
Thus, 
\begin{equation} 
{\rm PE}_{\ \!\rm ent} = \frac{1}{2} - \frac{1}{2} 
\sqrt{[1 - ({c_{2\theta}}^2 + s_{2 \phi} {s_{2 \theta}}^2)^2]}
\end{equation} 
for the two initial entangled states. 
This probability of error is minimal, hence it 
cannot be reduced by any physical process. Therefore, if, for some
$\theta$ and $\phi$, the disentanglement into the tensor product states 
{\em reduces} the ${\rm PE}$, then that process is non-physical.

The difference of the states obtained after disentangling into
tensor product states is
$ \Delta_{\rm disent} = \hat{\rho_0} \hat{\rho_0}-\hat{\rho_1} \hat{\rho_1}$
This matrix is
\begin{equation} 
\Delta_{\rm disent} =
\cos 2 \phi \sin 2 \theta 
\left( \begin{array}{cccc}
 0 & a & a & 0 \\
 a & 0 & 0 & b \\ 
 a & 0 & 0 & b \\ 
 0 & b & b & 0 
\end{array} \right) 
\ , 
\end{equation}
with 
$a = \cos^2 \phi \cos^2 \theta + \sin^2 \phi \sin^2 \theta$ and
$b = \cos^2 \phi \sin^2 \theta + \sin^2 \phi \cos^2 \theta$.    
After diagonalization, we can calculate the Trace-Norm,
so finally we get
\begin{eqnarray} 
{\rm PE}_{\rm \ \!disent} &=& \frac{1}{2} - 
\frac{1}{\sqrt2}\sin 2 \theta \cos 2 \phi \sqrt{a^2 + b^2} \nonumber \\
 &=& \frac{1}{2} - 
\frac{1}{2} s_{ 2 \theta} c_{ 2 \phi} 
\sqrt{1 + {c_{2 \phi}}^2 {c_{2 \theta}}^2 } 
\ .
\end{eqnarray} 

We can now observe that there are values of $\theta$ and $\phi$,
e.g., $\theta = \phi = \pi/8$,
for which the outcomes of the  disentanglement process are illegitimate
since they satisfy 
${\rm PE}_{\rm disent} < {\rm PE}_{\rm ent} $. 
Once these outcomes are illegitimate the disentanglement process 
leading to these outcomes is non-physical,
proving that a disentangling machine which disentangle 
the states $|\psi_0\rangle$ and
$|\psi_1\rangle $ cannot 
exist for these values of $\theta $ and $\phi$.
Therefore, this analysis provides a proof 
(similar to Terno's proof~\cite{Terno98})
that a universal machine performing 
disentanglement into tensor product states 
cannot exist. 

The following set of states can easily be disentangled:
\begin{equation} 
|\psi_0\rangle = \frac{1}{\sqrt2} [ |00\rangle + |11\rangle ]
\ ; \quad
|\psi_1\rangle = \frac{1}{\sqrt2} [ |00\rangle - |11\rangle ]
\end{equation}
To disentangle them, Eve's machine uses an ancilla which is
another pair of particles in a
maximally entangled state (e.g., the singlet state) in any basis. 
Eve's machine swaps one
of the above particles with one of the members of the added pair, and traces
out the ancillary particles.
As a result, the state of the remaining two particles (one from each
entangled pair) is 
\begin{equation}\label{cms}
(1/4)[
|00 \rangle \langle 00|  +
|01 \rangle \langle 01|  +
|10 \rangle \langle 10|  +
|11 \rangle \langle 11| ]
\ ,
\end{equation}
the completely mixed state in four dimensions.
This set provides a trivial example of the ability to perform the 
disentanglement process. It is a trivial case of disentanglement,
since the two states are orthogonal:
they can first be measured and distinguished, and then, once the state is
known, clearly it can be disentangled.

However, exactly the same disentanglement process can be used to
successfully disentangle a non-trivial set of states.
Let the basis used for the two states be a different basis (and not the
same basis), so the first state is still $|\psi_0\rangle$, 
and the second state is
\begin{equation} 
|\psi_1'\rangle = \frac{1}{\sqrt2} [ |0'0'\rangle - |1'1'\rangle ]
\ .
\end{equation}
The same process of disentanglement still works, while now the states are
non-orthogonal, and cannot always be successfully distinguished.
Hence, this disentanglement process is non-trivial.
Note that the same process successfuly works also when more than two 
maximally entangled 
states are used as the possible inputs.

Before we continue, let us recall some proofs of the no-cloning
argument, since the methods we use here are quite similar the those used in
the no-cloning argument.
Let the cloner obtain an unknown state and try to clone it.
To prove that this is impossible, it is enough to provide one set of states
for which the cloner 
cannot clone any state in this set. Let the sender (Alice) and
the cloner (Eve) use 
three states $|0\rangle$, $|1\rangle$, and 
$|+\rangle = (1/\sqrt2)[|0\rangle + |1\rangle]$.
The most general process which can be used here in the attempt of cloning
the unknown state from this set is
to attach an ancilla in an arbitrary dimension and in a
known state (say $|E\rangle$), to tranform the
entire system using an arbitrary unitary transformation, 
and to trace out the unrequired parts of the ancilla.
In order to clone the states 
$|0 \rangle$ and $|1\rangle$ the transformations are restricted
to be
\begin{equation}
|E0\rangle \longrightarrow |E_0 00\rangle 
\ ; \quad 
|E1\rangle \longrightarrow |E_1 11\rangle 
\end{equation}
and once the remaining ancilla is traced out, the cloning process is
completed.
Due to linearity, 
this fully determine the transformation of the last state to be
\begin{equation}
|E+\rangle] \longrightarrow 
\frac{1}{\sqrt2} [|E_0 00\rangle + |E_1 11\rangle 
\ ,
\end{equation}
while a cloning process should yield 
\begin{equation}
|E+\rangle \longrightarrow |E_+\! +\!+\rangle 
\ .
\end{equation}
The contradiction is clearly apparent since, once the remaining ancilla is
traced out, the second expression has a non-zero
amplitude for the term $|01\rangle $ while the first
expression does not.
The conventional way~\cite{WZ82}
of proving the no-cloning theorem (using only two
states, say $|0\rangle$ and $|+\rangle$) is to compare the overlap before and
after the transformation (it must be equal due to the unitarity of quantum
mechanics):
We obtain that 
$\langle E|E\rangle \langle 0 | + \rangle =
\langle E_0|E_+\rangle \langle 0 | + \rangle \langle 0 | + \rangle$.
Hence 
$ 1 = \langle E_0|E_+\rangle \langle 0 | + \rangle $ which
is obviously wrong since all the terms on the right hand side are smaller than
one.

We shall now use the linearity of quantum mechanics to
show that there are states that cannot be disentangled into
tensor product states, but can only be disentangled 
into a mixture of tensor
product states.
Surprisingly, our proof is
{\em mainly} based on the {\em disentanglement of product states},
that is, on the disentanglement of states 
which are anyhow not entangled even before the disentanglement process.
The reason for the usefulness of such states is that they provide strict 
restrictions on the allowed transformations.

The following set of states cannot be disentangled into product states:
\begin{eqnarray}
|\psi_0 \rangle &=& |00\rangle  \nonumber \\
|\psi_1 \rangle &=& |11\rangle  \nonumber \\
|\psi_2 \rangle &=& |00\rangle + |11\rangle 
\end{eqnarray}
We shall assume that these states can be disentangled into product states 
and we shall reach a contradiction.
Note that the resulting states should be 
$|\psi_0 \rangle$ and $|\psi_1\rangle $
in the first two cases (see Eq.~\ref{pure-ps}), and the resulting state
should be 
the completely mixed state (in 4 dimensions) in the last case
(see Eq.~\ref{cms}).

The most general process which can be used here is
to attach an ancilla in an arbitrary dimension and in a
known state (say $|E\rangle$), to transform the
entire system using an arbitrary unitary transformation, 
and to trace out the ancilla.
In order to avoid changing the states 
$|\psi_0 \rangle$ and $|\psi_1\rangle$ the transformations are restricted
to be
\begin{eqnarray} \label{two-states}
|E\psi_0\rangle &=& |E00\rangle \longrightarrow |E_0 00\rangle \nonumber
\\
|E\psi_1\rangle &=& |E11\rangle \longrightarrow |E_1 11\rangle 
\ .
\end{eqnarray}
As in the no-cloning argument, these transformations
fully determine the transformation of the last state to be
\begin{equation}
|E\psi_2\rangle \longrightarrow \frac{1}{\sqrt2} [|E_0 00\rangle 
+ |E_1 11\rangle 
\ .
\end{equation}
Once we trace out the ancilla, the resulting state
is still entangled  
unless $|E_0\rangle$ and $E_1\rangle$
are orthogonal. The proof of that statement is as follows:
Without loss of generality the states $|E_0\rangle $ and $|E_1\rangle $
can be
written as $|E_0 \rangle = |0 \rangle$ and 
$|E_1 \rangle =  \alpha |0 \rangle + \beta |1 \rangle$ with
$|\alpha^2| + |\beta^2| = 1$.
Thus, 
$|E\psi_2\rangle \longrightarrow \frac{1}{\sqrt2} |0\rangle ( |00\rangle 
+ \alpha |11\rangle  ) +  \frac{\beta}{\sqrt2}  |111\rangle$.
When the ancilla is traced out the remaining state is 
\begin{equation} \label{resulting} 
\left( \begin{array}{cccc}
 1/2     & 0       & 0       & \alpha^*/2 \\
 0       & 0       & 0       & 0       \\ 
 0       & 0       & 0       & 0       \\ 
\alpha/2 & 0       & 0       &1/2
\end{array} \right) 
\ . 
\end{equation}
The resulting state is entangled unless 
$\alpha = 0$. Thus, in a successful disentanglemet process 
$\alpha = 0$ and hence, $|E_1\rangle = e^{i\theta} |1\rangle$. 
This state, however, is not a tensor product state, thus the above set of
states {\em cannot} be disentangled into tensor product states.

At the same time, this example also shows that the above set 
of states {\em can} be
disentangled into a mixture of tensor product states. 
The resulting state (\ref{resulting})   
still has the correct reduced density matrices
for each subsystem---the completely mixed state in two dimensions.
With $\alpha=0$, the resulting state is  
$(1/2) [|00\rangle \langle 00| + |11\rangle \langle 11|] $,
so we succeeded in showing an example where the states can only be
disentangled into a separable state, but not into a tensor product state.

Our result resembles a result regarding two commuting
mixed states~\cite{broadcast}: these states cannot be cloned, but they can be
broadcast. That is, the resulting state of the cloning device cannot be a
tensor product of states which are equal to the original states, but can be
a separable state whose reduced density matrices are equal to the original
states~\cite{Bennett}.

At that stage, the main question (raised by~\cite{Terno98}  
and~\cite{Fuchs}) 
is still left open: Can there be a universal
disentangling machine?
That is, can there exist a machine that disentangles
any set of states into separable states?
We shall now show that such a machine cannot exist.

Our result is obtained by combining several of 
the previous techniques: the use of 
linearity, unitarity, and the 
disentanglement of product state. 

Consider the following set of states
\begin{eqnarray}
|\psi_0 \rangle &=& |00\rangle  \nonumber \\
|\psi_1 \rangle &=& |11\rangle  \nonumber \\
|\psi_2 \rangle &=& (1/\sqrt2)[ |00\rangle + |11\rangle ] \nonumber \\
|\psi_3 \rangle &=& |++\rangle  \nonumber \\
\end{eqnarray}
in which we added the states
$|\psi_3 \rangle$  to the previous set.
This set of states cannot be disentangled.

The allowed transformations are now more restricted since, 
in addition to (Eq.~\ref{two-states}),
the state
$|\psi_3\rangle$ 
must also not be changed by the disentangling machine:
\begin{equation}
|E \psi_3\rangle = |E\!+\!+\rangle \longrightarrow |E_+\! +\!+\rangle  
\ .
\end{equation}
Due to unitarity, 
$\langle E|E\rangle \langle 0 | + \rangle \langle 0 | + \rangle =
\langle E_0|E_+\rangle \langle 0 | + \rangle \langle 0 | + \rangle$, and
also 
$\langle E|E\rangle \langle 1 | + \rangle \langle 1 | + \rangle =
 \langle E_1|E_+\rangle \langle 1| + \rangle \langle 1| + \rangle$. 
These expressions yield 
$ 1 = \langle E_0|E_+\rangle $,
and
$ 1 = \langle E_1|E_+\rangle $,
from which we must conclude that 
$|E_+\rangle = |E_0\rangle = |E_1\rangle$.
Recall that we already found that $|E_0\rangle = |0 \rangle$ and 
$|E_1\rangle = e^{i\theta} |1 \rangle$, due to the disentanglement of
$|\psi_2\rangle$.    
Since the two requirements contradict each other, the proof that the above
set of states cannot be disentangled (not even to a separable state)
is completed.
Thus, we have proved that a universal disentangling machine cannot exist.
In otehr words---a single quantum cannot be disentangled.

This result resembles a result regarding two non-commuting
mixed states~\cite{broadcast}: these states cannot be cloned, and
furthermore, they cannot be
broadcast. 

To summarize, we provided a thorough analysis 
of disentanglement processes, and 
we proved that a single quantum cannot be disentangle.
Interestingly, we used a set of four states to prove this, but we conjecture
that there are smaller sets that could be used to establish the same
conclusion.

The no-cloning of states of composite systems were investigated 
recently~\cite{BDFMRSSW,Mor98}, and it seems that several interesting
connections between these works and the idea of disentanglement can 
be further explored.   
For instance, one can probably find systems where the states can
only be disentangled (or only be disentangled into product states) if the
two subsystems are available together, but {\em cannot} be disentangled if the
subsystems are available one at a time 
(with similarity to~\cite{Mor98}),
or {\em cannot} be disentangled if only 
bilocal superoperators 
can be used for the disentanglement process
(with similarity to~\cite{BDFMRSSW}).

I would like to thank Charles Bennett, Oscar Boykin, 
and Danny Terno for very helpful remarks and discussions.

\end{multicols}

\end{document}